# Dynamic magnetic-transformation-induced exchange bias in ($\alpha$-Fe$_2$O$_3$)$_{0.1}$-(FeTiO$_3$)$_{0.9}$

P. Song[1,2], L. Ma[1,a)], G. K. Li[1], C. M. Zhen[1], C. Wang[2], W. H. Wang[3], E. K. Liu[3], J. L. Chen[3], G. H. Wu[3], Y. H. Xia[4], J. Zhang[4], C. M. Xie[4], H. Li[4] and D. L. Hou[1,b)]

[1]Department of Physics, Hebei Advanced Thin Films Laboratory, Hebei Normal University, Shijiazhuang 050024, China.
[2]Center for Condensed Matter and Material Physics, Department of Physics, Beihang University, Beijing 100191, China.
[3]Beijing National Laboratory for Condensed Matter Physics, Institute of Physics, Chinese Academy of Sciences, Beijing 100190, China.
[4]Key Laboratory of Neutron Physics, Institute of Nuclear Physics and Chemistry, China Academy of Engineering Physics, Mianyang 621999, China.

Up to now, for the conventional exchange bias (EB) systems there has been one pinning phase and one pinned phase, and the pinning and pinned phases are inherent to the material and do not mutually transform into each other. Interestingly, we show here that EB is observed in a special system ($\alpha$-Fe$_2$O$_3$)$_{0.1}$–(FeTiO$_3$)$_{0.9}$ (HI-9) different from the conventional EB system. Neutron powder diffraction and magnetic measurement confirm that for HI-9: i) two types of short-range antiferromagnetic ordering coexist; ii) there are two pinning phases and one pinned phase; iii) the pinned phase is not intrinsic to the structure but can be dynamically produced from the pinning phase with the help of an external magnetic field. Consequently, two anomalous EB behaviors are observed: i) both the coercivity ($H_C$) and the exchange bias field ($H_E$) simultaneously decrease to zero at 30 K; ii) for a high cooling field ($H_{cool}$), $H_E$ decreases logarithmically with increasing $H_{cool}$. Using Arrott plots it is confirmed that the first-order magnetic phase transformation (FOMPT) from the AFM Fe$^{2+}$ to ferromagnetic (FM) Fe$^{2+}$ and the second-order magnetic phase transformation (SOMPT) for the process whereby the FM Fe$^{2+}$ aligns with the external field direction coexist in HI-9. The Morin transition and FOMPT cause the anomalous EB behaviors. This work may provide fresh ideas for research into EB behavior.

Keywords: exchange bias, field-induced dynamic magnetic transformation, first-order magnetic phase transformation

---

a) corresponding author. majimei@126.com
b) corresponding author. houdenglu@mail.hebtu.edu.cn

# Introduction

After its discovery in the Co/CoO system, exchange bias (EB) behavior has been observed in many different materials, including nanoparticles,[1-5] inhomogeneous materials,[6,7] coated antiferromagnetic single crystals[8,9] and thin films.[10-12] In these EB materials, there usually exist two magnetic phases. One phase with strong magnetic anisotropy serves as the pinning phase and the other phase with weak magnetic anisotropy serves as the pinned phase. This is the case for EB in antiferromagnetic (AFM)-ferromagnetic (FM) systems,[13-16] ferrimagnetic (FIM)-FM systems,[17,18] AFM-FIM systems,[19] AFM-superferromagnetic (SFM)[20] or spin-glass (SG)-FM systems.[21,22] These EB systems form a "one pinning phase and one pinned phase" structure.[23] Illustration (I) in **Figure** 1(d) shows this EB structure. A typical feature of this conventional structure is that the pinning and pinned phases are inherent to the material, and do not mutually transform into each other. Interestingly, our group has recently observed EB phenomena in ($\alpha$-Fe$_2$O$_3$)$_{0.1}$–(FeTiO$_3$)$_{0.9}$ (referred to below as HI-9).[6] It is known that both $\alpha$-Fe$_2$O$_3$ and FeTiO$_3$ have a long-range AFM ordering and their Neel temperatures ($T_N$) are ~ 260 K and ~ 50 K, respectively.[24,25] The illustration of **Figure** 1b shows the crystal and magnetic structures of HI-9, both the Fe$^{3+}$ cations (red) between the neighboring layers and the Fe$^{2+}$ layers (orange) separated by the nonmagnetic Ti$^{4+}$ layers (gray) are AFM-coupled and their spins are parallel to the $c$-axis. Hereby, a system with two types of short-range AFM ordering forms when a very small amount of $\alpha$-Fe$_2$O$_3$ is doped into FeTiO$_3$ (no more than 20 %).[6,26] Considering that EB in AFM-AFM systems is rarely reported, HI-9 will be a good platform to study EB.

In this paper, two types of short-range AFM ordering, a "two pinning phases and one pinned phase" (TPPOPP) structure, and the field-induced dynamic magnetic transformation between the pinning and pinned phases are evidenced by the neutron powder diffraction and magnetic measurement. Two anomalous EB behaviors are observed: i) both the coercivity ($H_C$) and the exchange bias field ($H_E$) simultaneously decrease to zero at 30 K; ii) for a high cooling field ($H_{cool}$), $H_E$ decreases logarithmically with increasing $H_{cool}$. By application of Arrott plots, it has been confirmed that the above phenomena are related to the Morin transition and the first-order magnetic phase transformation (FOMPT).

**Experimental details**

The HI-9 sample was prepared using a solid-state reaction method. Ground powders of $FeTiO_3$ (99.98 %) and $\alpha$-$Fe_2O_3$ (99.99 %) were fully mixed and sintered at 1473 K for 12 h and cooled slowly to room temperature. The neutron powder diffraction (NPD) experiments were carried out at various temperatures using high resolution neutron powder diffractometer (HRND) ($\lambda$=1.884 Å) at China Mianyang Research Reactor (CMRR). Powder X–Ray diffraction (XRD) patterns with Cu-$K\alpha$ radiation were obtained ranging from 20 to 300 K. Crystal structures were determined by the Rietveld refinement method with the General Structure Analysis System (GSAS).[27] To measure the magnetic properties of the powders, the sample was compressed into a nonmagnetic capsule. The magnetic measurements were performed in the temperature range from 2 to 300 K using a Physical Property Measurement System (PPMS-9, Quantum Design).

**Results and discussion**

1. **Field-induced dynamic magnetic transformation of the TPPOPP structure in the AFM-AFM coupled system of HI-9.**

Figure 1a shows the neutron powder diffraction (NPD) patterns and magnetic hysteresis (M-H) loop of HI-9 at room temperature (300 K) and low temperature (5 K), respectively. Rietveld analysis including occupancy refinement was performed to fit the NPD data. The fitted results confirm that HI-9 does have the crystal structure as shown in Figure 1b, and the occupancy of the $Fe^{3+}$ cations in the $Fe^{2+}$ and $Ti^{4+}$ layers is 0.091 and 0.109, respectively. Most importantly, it is found that there is no contribution from magnetic scattering to the Bragg peaks at either room temperature or low temperature, suggesting that there is no long-range magnetic ordering throughout the entire temperature range, which is consistent with the short-range AFM ordering of the system.[28,29,30] However, the magnetic hysteresis loops of HI-9 at room temperature and low temperature are very different, unlike the case of NPD. There is no magnetic hysteresis at 300 K, but the obvious hysteresis is observed at 5 K and the saturation magnetization is as high as 2.22 $\mu_B$/Fe, suggesting the existence of the long-range FM ordering.

**Figure** 1b shows the temperature dependence of the normalized relative lattice constants ($c$-$c_{300k}$)/$c_{300k}$ =$\Delta c$/$c_{300k}$ and ($a$-$a_{300k}$)/$a_{300k}$=$\Delta a$/$a_{300k}$ based on the refinements of the variable-temperature NPD and XRD data. It can be seen that $\Delta a$/$a_{300K}$ almost linearly decreases with decreasing temperature, thus exhibiting a normal positive thermal expansion behavior, while $\Delta c$/$c_{300K}$ also decreases linearly initially, but an abnormal negative expansion occurs for temperature lower than ~ 100 K, and this behavior becomes more pronounced for temperature lower than ~ 30 K. According to our previous work6 and Burton *et al*.'s work,[31] ~ 100 K corresponds to the magnetic transformation temperature of FeTiO$_3$ phase from the paramagnetic (PM) phase to the PM' state and which is a subfield of the PM phase. On the other hand, based on the results from parts 1 and 2 of the supplementary material, ~ 30 K corresponds to the Morin transition temperature ($T_M$) of the $\alpha$-Fe$_2$O$_3$ phase, *i. e.*, the Fe$^{3+}$ cations will transform from the canted-AFM perpendicular to the $c$ axis to the collinear-AFM parallel to the $c$ axis as the temperature is lower than 30 K.[24] Therefore, the abnormal negative expansions in $\Delta c$/$c_{300K}$ are ascribed to the spontaneous magnetostriction induced by these two types of the spontaneous magnetization.[32] As a result, two types of short-range AFM ordering are obtained in HI-9 when the temperature is lower than 30 K. The first is AFM ordering of the Fe$^{3+}$ cations (red) between the neighboring layers, and the second is AFM ordering of the Fe$^{2+}$ cations (orange) between the alternating layers, as shown in **Figure** 1b.

**Figure** 1c shows the maximum applied magnetic field ($H_m$) dependence of the hysteresis loop at 5 K, and the inset displays the partial enlarged view. It is observed that the loops are reversibly linear when $H_m$ is lower than 5 kOe, which is consistent with the fact that two types of short-range AFM ordering coexist at 5 K, but the loops begin to show the irreversible hysteresis when $H_m$ is equal to or larger than 5 kOe, which is a typical FM behavior. On the other hand, the rapid linear increase in magnetization from 30 kOe to 40 kOe is also observed, suggesting the existence of AFM ordering. The above results confirm that for one of two types of short-range AFM ordering, a field-induced dynamic magnetic transformation from AFM to FM has occurred (the critical field is ~ 5 kOe), and the other type of short-range AFM ordering maintains the same. The former corresponds to the $Fe^{2+}$ cations, and the latter corresponds to the $Fe^{3+}$ cations.[6,33]

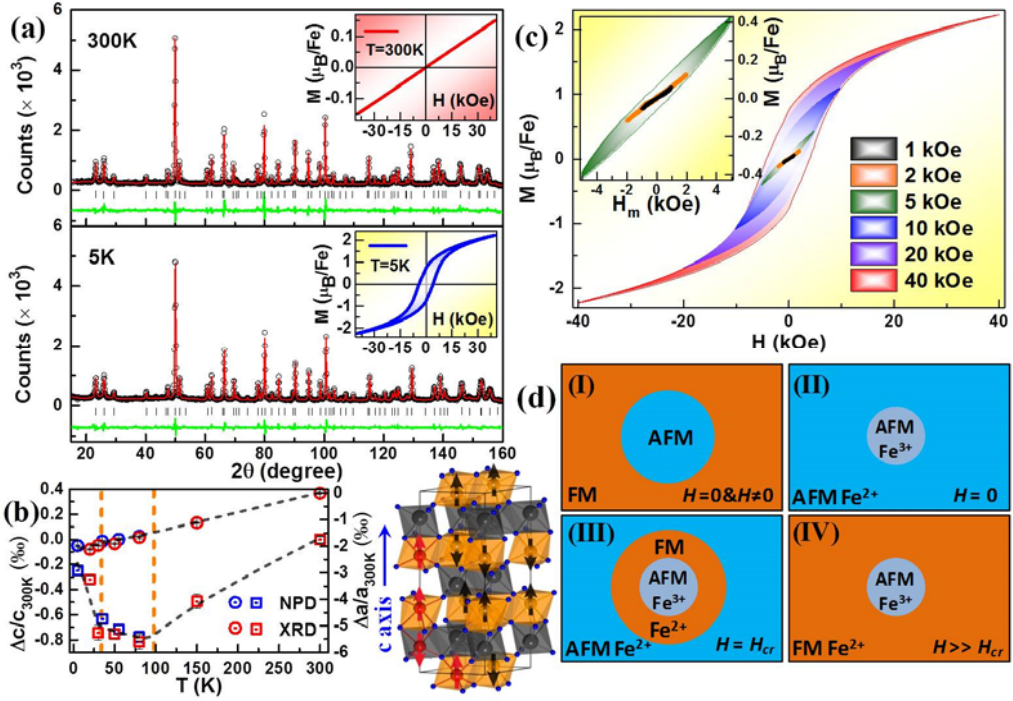

**Figure 1 Three characteristics in HI-9. (a)** Neutron powder diffraction patterns of HI-9 at 300 K and 5 K. The circles show the experimental intensities ($I_{obs}$), the red solid line shows the calculated intensities ($I_{calc}$), the green solid line is the difference between the observed and calculated intensities ($I_{obs} − I_{calc}$), and the vertical lines indicate the angular positions of the Bragg reflections. Insets are the magnetic hysteresis loops at 300 K and 5 K, respectively. **(b)** Temperature dependence of the normalized relative lattice constants ($c-c_{300k})/c_{300k} =\Delta c/c_{300k}$ and ($a-a_{300k})/a_{300k}=\Delta a/a_{300k}$ based on the refinements of the variable-temperature NPD and XRD data, and the dash lines are to guide the eye. Right illustration is the crystal and magnetic structures of HI-9. **(c)** Maximum applied magnetic field ($H_m$) dependence of the magnetic hysteresis loop at 5 K, and $H_m$ is changed from 1 kOe to 40 kOe. Inset displays the partial enlarged view. **(d)** Schematic diagram of the magnetic phase in the conventional exchange bias system (I) and in the HI-9 system (II-IV) under different external fields. AFM, FM, $H$ and $H_{cr}$ represent antiferromagnetic, ferromagnetic, external magnetic field and the critical external magnetic field, respectively.

Illustration II-IV in **Figure** 1d show the schematic diagram of the magnetic phase in HI-9 under different fields, based on the above experimental results. First, two types of short-range AFM ordering coexist when the external field is zero, and they together serve as the pinning phase (see illustration II). Then part of AFM $Fe^{2+}$ transform into FM state when the external field increases to the critical field (see illustration III), and the formed FM $Fe^{2+}$ serves as the pinned phase. Finally, all of AFM $Fe^{2+}$ transform into FM state when the external field is large enough (see illustration IV), and the pinning phase of AFM $Fe^{2+}$ totally vanishes and the pinned phase of FM $Fe^{2+}$ reaches the maximum. In brief, for HI-9 the pinned phase is not intrinsic but can be dynamically transformed from the pinning phase with the help of the external field. Therefore, compared to the conventional EB system (see illustration I), HI-9 can serve as a special EB platform, due to its three characteristics: i) its belonging to an AFM-AFM system, ii) its "two pinning phases and one pinned phase" (TPPOPP) structure, and iii) the field-induced dynamic magnetic transformation between the pinning and pinned phases.

**2. Two observed anomalous EB behaviors in HI-9.**

Figure 2 a, b and c show the temperature dependence of the coercivity ($H_C$), the exchange bias field ($H_E$) and the saturation magnetization ($M_S$) in HI-9, for different values of the cooling field $H_{cool}$. It is interesting to observe that both $H_C$ and $H_E$ decrease rapidly with increasing temperature, and they simultaneously reach zero at 30 K. As is known to us, the conventional EB systems usually show two other different cases based on the relative anisotropy strength between the pinning and the pinned phases. If the pinning phase has a relatively weaker anisotropy, $H_C$ usually reaches its maximum when $H_E$ is reduced to zero, in such case both the pining and the pinned phases contribute to $H_C$;[13,34] whereas if the pinning phase has a relatively stronger anisotropy, there is almost no change in $H_C$ when $H_E$ is reduced to zero, in such case only the pinned phase contributes to $H_C$.[5,35] Therefore, the correlation between $H_C$ and $H_E$ in HI-9 is completely different from that of the conventional EB systems, and it is regarded as the first anomalous EB behavior of HI-9.

Why does $H_C$ become zero at the blocking temperature of EB or the Morin transformation temperature (30 K)? It is known that $H_C$ originates from a variety of magnetic anisotropies in magnetic materials. In the conventional EB system, $H_C$ usually originates from three types of anisotropies: i) the intrinsic anisotropy of the pinned phase, ii) the unidirectional acquired anisotropy of the pinned phase at the interfaces, and iii) the anisotropy of the pinning phase dragged by the pinned phase. In the HI-9 system, only the first two anisotropies exist due to the fact that $Fe^{3+}$ cations (the pinning phase) cannot be dragged by $Fe^{2+}$ cations (the pinned phase). Considering that the intrinsic anisotropy of $Fe^{2+}$ is not inherent but rather induced by the dynamic magnetic transformation from AFM to FM under the constraints of interface pinning of the AFM $Fe^{3+}$, thus the first two anisotropies are coupled together and coexist. When the temperature is higher than 30 K, $Fe^{3+}$ will transform from the collinear-AFM parallel to the *c* axis to the canted-AFM perpendicular to the *c* axis,[24] i.e., the interface pinning interaction disappears. As a result, the first two anisotropies in HI-9 will vanish simultaneously, accordingly both $H_C$ and $H_E$ will reach zero at 30 K simultaneously. Compared to the case of $H_C$ and $H_E$, the case of $M_S$ is different. When the temperature is lower than 30 K, $M_S$ is almost unchanged (~2.25 μ$_B$/Fe); whereas $M_S$ dramatically decreases, when the temperature is higher than 30 K as seen in **Figure** 2c. It is noteworthy that the temperature dependence of $M_S$ is completely different from that of $H_C$ and $H_E$, but the physical mechanism for generating these temperature dependences are the same. They are all due to the disappearance of the collinear AFM-AFM interaction between $Fe^{3+}$ and $Fe^{2+}$ cations and the dynamic magnetic transformation from AFM $Fe^{2+}$ to FM $Fe^{2+}$. Therefore, it is the collinear AFM-AFM system and the dynamic magnetic transformation that results in the first anomalous EB behavior of HI-9.

**Figure** 2d and e show the $H_{cool}$ dependence of $H_E$ at different temperatures (*T* = 2.5 K, 3 K and 5 K) in HI-9. With increasing $H_{cool}$, $H_E$ first increases rapidly when $H_{cool}$ is lower than 10 kOe, then it falls off logarithmically when $H_{cool}$ is higher than 10 kOe as shown in **Figure** 2e where the horizontal axis is ln($H_{cool}$). The $H_{cool}$ dependence of $H_E$ in HI-9 is also different from that of the conventional EB materials in which $H_E$ either saturates[13] or falls off linearly not logarithmically with further increasing $H_{cool}$.[21,36] The logarithmic decrease of $H_E$ is thus regarded as the second anomalous EB behavior in HI-9.

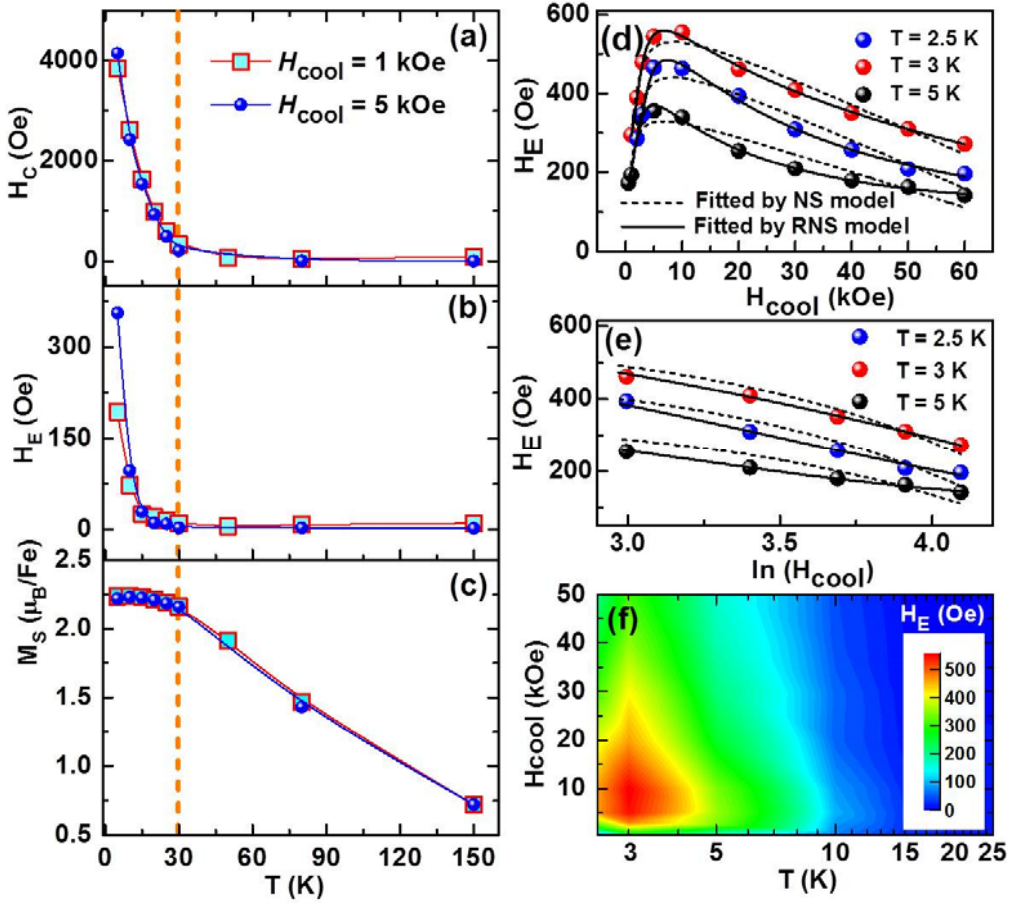

**Figure 2 Two abnormal exchange bias behaviors in HI-9.** Temperature dependence of $H_C$ **(a)**, $H_E$ **(b)** and $M_S$ **(c)** with different cooling fields ($H_{cool}$ = 1 and 5 kOe). The dotted line shows the position of the $T$ = 30 K critical temperature. Cooling field ($H_{cool}$) **(d)** and ln ($H_{cool}$) **(e)** dependence of $H_E$ at different temperatures ($T$ = 2.5 K, 3 K and 5 K). Dotted lines and solid lines show the fitting results according to the NS model and the RNS model, respectively. **(f)** Contour map of temperature and cooling field co-dependence of $H_E$.

For the linear decrease of $H_E$ in the conventional EB materials, Niebieskikwiat and Salamon[23] have put forward a simplified exchange interaction model (referred to below as the NS model) as,

$$-H_E \propto J_i \left[ \frac{J_i \mu_0}{(g\mu_B)^2} L\left(\frac{\mu H_{cool}}{k_B T_f}\right) + H_{cool} \right] \quad (1)$$

where $J_i$ is the interface exchange constant, $g$ is the Lande factor, $\mu_B$ is the Bohr magneton, $L(x)$ is the Langevin function, $\mu = N_v \mu_0$, $N_v$ is the number of FM spins, $\mu_0$ is the vacuum permeability and $k_B$ is the Boltzmann constant. The dotted lines in **Figure 2d** and **e** show the fitting results according to the NS model. It is seen that there is some deviation between the experimental results and the NS model, maybe due to the field-induced dynamic magnetic transformation in HI-9 as mentioned above. Now the field-induced dynamic magnetic transformation based on the TPPOPP structure is taken into account to modify the NS model. The revised NS model (referred to below as the RNS model) can be expressed as,

$$-H_E \propto J_i \left[ \frac{J_i \mu_0}{(g\mu_B)^2} L\left(\frac{\mu H_{cool}}{k_B T_f}\right) + H_{cool} - \ln(H_{cool}) \right] \quad (2)$$

where a new term $-J_i \ln(H_{cool})$ appears compared to the NS model. The detailed derivation of the RNS model is given in part 3 of the supplementary material. The solid lines in **Figure** 2d and e are the fitting results according to the RNS model. Obviously, the RNS model is in good agreement with the experimental data. It is confirmed that it is the TPPOPP structure and the field-induced dynamic magnetic transformation that lead to the second anomalous EB behavior in HI-9. **Figure** 2f displays the contour map of temperature and cooling field co-dependence of $H_E$. $H_E$ is extremely sensitive to $H_{cool}$. $H_E$ will reach its maximum under appropriate temperature and cooling field. This may be correlated to the field-induced dynamic magnetic transformation.

## 3. Coexistence of first-order magnetic phase transformation (FOMPT) and second-order magnetic phase transformation (SOMPT) in HI-9.

Both the TPPOPP structure and the two abnormal EB behaviors are correlated to the field-induced dynamic magnetic transformation. Now Arrott plots, in which the initial magnetization data is presented in the form of $M^2$ versus $H/M$ isotherms, are applied to investigate the field-induced dynamic magnetic transformation in HI-9. If Arrott plots have a negative slope, the transformation is the first-order magnetic phase transformation (FOMPT); if Arrott plots show a positive slope, the transformation is the second-order magnetic phase transformation (SOMPT).[37-39] **Figure** 3a shows Arrott plots of HI-9 obtained from the initial magnetization curves from 5 K to 40 K which are shown in part 4 of the supplementary material. Obviously, except for the 40 K curve all Arrott plots have both negative and positive slopes, and for each curve there exists a critical magnetic field ($H_{cmf}$) represented by the yellow solid circles in **Figure** 3a, at which the slope changes from negative to positive. It is found that for $H < H_{cmf}$ the slope is negative, i. e., FOMPT occurs; and for $H > H_{cmf}$ the slope is positive, i. e., SOMPT occurs. FOMPT in the low external field should correspond to the field-induced dynamic magnetic transformation from AFM $Fe^{2+}$ to FM $Fe^{2+}$ (see the left and middle schematic representations in **Figure** 3b). Charilaou *et al.*[33] have reported that heat is released during the magnetization jump of $Fe^{2+}$ from AFM to FM. Our result also demonstrates that the transformation of $Fe^{2+}$ in HI-9 is first order. SOMPT in the high external field may correspond to the process of FM $Fe^{2+}$ aligning with the external field direction (see the middle and right schematic representations in **Figure** 3b). Therefore, FOMPT and SOMPT coexist in HI-9 when the temperature is lower than 30 K, and only SOMPT exists in HI-9 when the temperature is higher than 30 K.

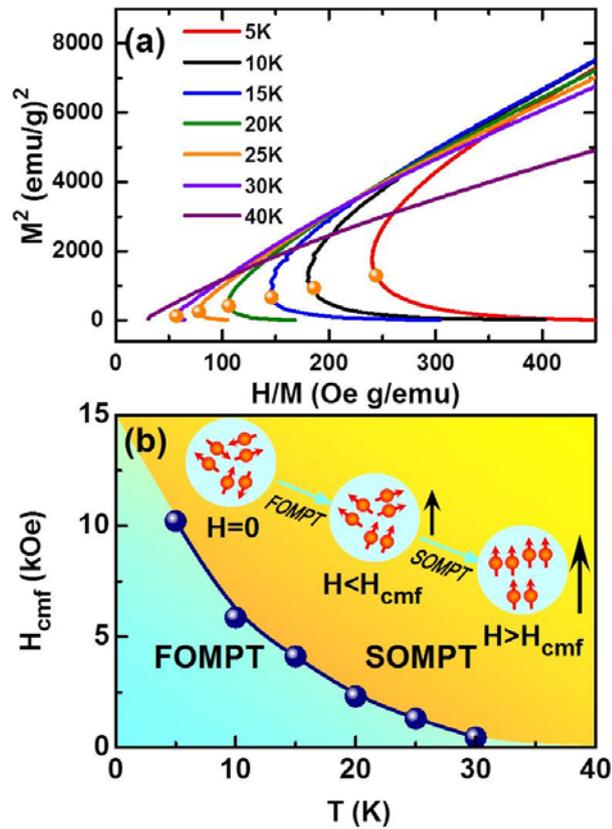

**Figure 3 Field-induced dynamic magnetic transformation in HI-9.** (a) Arrott plots from 5 K to 40 K. The yellow spheres represent the position of the critical magnetic field ($H_{cmf}$) where the curve slope changes from negative to positive. (b) Temperature dependence of $H_{cmf}$. The solid line is to guide the eye. The inset shows the schematic representations of the evolution from the first-order magnetic phase transformation (FOMPT) to the second-order magnetic phase transformation (SOMPT) with increasing the external magnetic field. Red and black arrows mark the orientation

of the $Fe^{2+}$ moments and the external magnetic field, respectively.

**Figure** 3b shows the temperature dependence of $H_{cmf}$. It is seen that $H_{cmf}$ decreases dramatically with increasing temperature and approaches zero at 30 K. This indicates that the external field required to complete FOMPT become smaller and smaller, and finally FOMPT disappears for $T > 30$ K. It is noteworthy that with decreasing temperature all anomalies occur at 30 K, including the abnormal negative expansion in $c$ axis, the Morin transition of $Fe^{3+}$, exchange bias, coercivity, the constant saturation magnetization, and first-order magnetic transformation. Therefore, all of these are coupled together. The Morin transition of $Fe^{3+}$ leads to the abnormal negative expansion in $c$ axis, and the emergence of the pinning phase. FOMPT generates FM $Fe^{2+}$, i. e., the pinned phase. Consequently, exchange bias effect (non-zero $H_E$ and $H_C$) and the constant $M_S$ in HI-9 appear. We conclude, therefore, that it is the Morin transition and FOMPT that cause the anomalous EB behaviors.

## Conclusion

In summary, $(\alpha\text{-}Fe_2O_3)_{0.1}$–$(FeTiO_3)_{0.9}$ (HI-9) can serve as a special platform for investigating exchange bias (EB) behavior, due to its three characteristics: i) an antiferromagnetic (AFM)-AFM system, ii) the "two pinning phases and one pinned phase" structure and iii) the field-induced dynamic magnetic transformation between the pinning and pinned phases. These characteristics are supported by the results of the neutron powder diffraction and the magnetic dynamic measurement. Two anomalous EB behaviors are observed in HI-9: i) both the coercivity ($H_C$) and the exchange bias field ($H_E$) simultaneously decrease to zero at 30 K due to the collinear AFM-AFM system and the dynamic magnetic transformation, ii) for a high cooling field ($H_{cool}$) $H_E$ decreases logarithmically with increasing $H_{cool}$ due to the "two pinning phases and one pinned phase" structure and the field-induced dynamic magnetic transformation, based on which a revised Niebieskikwiat and Salamon model is put forward and it can explain well the logarithmical decrease in $H_E$. Arrott plots confirm that the first-order magnetic phase transformation (FOMPT) corresponding to the field-induced dynamic magnetic transformation from AFM $Fe^{2+}$ to FM $Fe^{2+}$ and the second-order magnetic phase transformation (SOMPT) for the process of FM $Fe^{2+}$ aligning with the external field direction coexist in HI-9. The Morin transition of $Fe^{3+}$ cations and FOMPT cause the anomalous EB behaviors in HI-9. This work may provide fresh ideas for research into EB behavior.

See supplementary material for the verification of the characteristic temperature, the detailed derivation of the RNS model, and the initial magnetization curves at different temperatures for HI-9.

## Acknowledgement

This work was supported by the Hebei Natural Science Foundation (Nos. A2014205051, E2016205268, and A2017210070), China Postdoctoral Science Foundation (No.2016M600192), the Key Project of Natural Science of Hebei Higher Education (No. ZD2017045), and the National Natural Science Foundation of China (No. 11504247). We thank Dr. Kewen Shi in Beihang University for help in the refinement of the crystal structure.

**Supplementary information:**

**1) Verification of the characteristic temperature using dc and ac susceptability**

**Figure** S1 shows the temperature dependence of the FC and ZFC dc susceptibilities ($\chi_{dc}$) under a field of $H = 50$ Oe (a) and the real part ($\chi'_{ac}$) (b) and imaginary part ($\chi''_{ac}$) (c) of the ac susceptibility at various frequencies for ($\alpha$-Fe$_2$O$_3$)$_{0.1}$–(FeTiO$_3$)$_{0.9}$. Three characteristic temperatures of these curves are worthy of note as suggested by the red dash lines.

The first characteristic temperature is 60 K and corresponds to the formation temperature of the partially disordered antiferromagnetic (PDA) state. J. Appl. Phys. 115, 213907 (2014) gives a detailed discussion of this characteristic temperature.

The second characteristic temperature is ~ 46 K and it is the freezing temperature of the cluster spin glass (CSG). To determine this characteristic temperature, three methods were adopted. Firstly, the parameter P = $\Delta T_f / (T_f \lg \omega)$ was used to describe the magnetic order. The value of P depends strongly on the interaction between the particles or the magnetic clusters. For systems with noninteracting entities such as in the case of superparamagnetism, the value of P is ~ 0.1. Any interactions weaken this value. For a CSG, the value of P is ~ 0.01; for a canonical SG, the value is 0.001 and for a well-ordered FM or antiferromagnetic system the value is almost zero. [Phys. Rev. B 76, 224419 (2007)] For HI-9, the value of P is 0.02, corresponding to a CSG. Secondly, the power law, $\tau = \tau^*_0 [(T_f - T_g) / T_g]^{-zv}$, [Phys. Rev. B 84, 024409 (2011)] was used. The fitting curve is shown in (d) of **Figure** S1 and the fitting parameter values are listed in **Table** S1. Thirdly, the empirical Vogel–Fulcher law, $\omega = \omega_0 \exp[-E_a / k_B (T_f - T_0)]$ [J. A. Mydosh, *Spin Glasses: An Experimental Introduction* (Taylor & Francis, London, 1993), p. 68.] was used. The fitting curve is shown in (e) of **Figure** S1 and the fitting parameter values are also listed in **Table** S1. All these three methods confirm that the second characteristic temperature (46 K) is the freezing temperature of the CSG.

The third characteristic temperature is 30 K. We confirmed that it is not the freezing temperature of spin glass or blocking temperature of superparamagnetism on the basis of two experimental facts: i) the $\chi'_{ac}$ and $\chi''_{ac}$, as shown in **Figure** S1b and c, have no peak at 30 K, ii) as shown in **Figure** S2 there is still an obvious magnetic relaxation in HI-9 up to 40 K. We also confirmed that 30 K is the Morin temperature of the α-Fe$_2$O$_3$ phase on the basis of three points: i) the dc FC susceptibility, as shown in **Figure** S1a, abruptly decreases at 30 K with decreasing temperature, ii) the Morin temperature is verified using a Clausius-Clapeyron type equation in part 3 of the supplementary material, iii) doping with transition elements, such as Al, Ti etc., can suppress the Morin transition of α-Fe$_2$O$_3$, for example Fe$_{2-x}$Ti$_x$O$_3$ does not exhibit the Morin transition down to 4.2 K, [A. H. Morrish, *Canted Antiferromagnetism: Hematite* (World Scientific, Singapore, 1994), p. 148; Nat. Nanotechnol. 2, 631 (2007)], thus the Morin temperature (30 K) of our case is entirely possible.

In conclusion, there exist three characteristic temperatures. 60 K is the formation temperature of the PDA state; 46 K is the freezing temperature of CSG, and 30 K is the Morin transition temperature of α-Fe$_2$O$_3$.

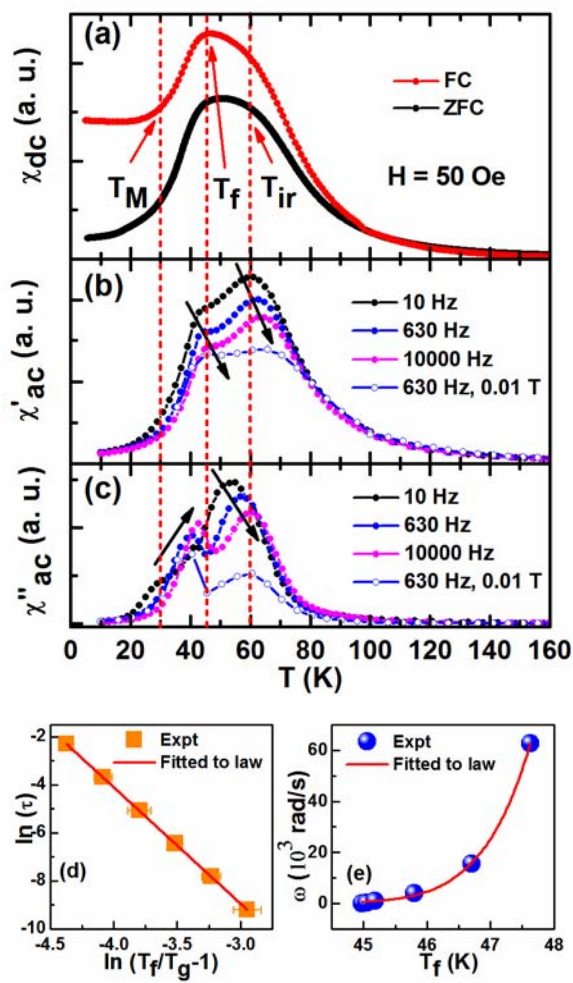

**Figure S1 (a)** FC (red solid circles) and ZFC (black solid circles) dc susceptibility as a function of temperature for $(FeTiO_3)_{0.9}$–$(Fe_2O_3)_{0.1}$ under $H = 50$ Oe. Real part ($\chi'_{ac}$) **(b)** and imaginary part ($\chi''_{ac}$) **(c)** of the ac susceptibility in an ac field of 10 Oe at various selected frequencies as a function of temperature for $(FeTiO_3)_{0.9}$–$(Fe_2O_3)_{0.1}$. $\chi'_{ac}$ and $\chi''_{ac}$ at 630 Hz (open circles) were measured under an ac field of 10 Oe and a dc field of 0.1 kOe. Black arrows indicate the changes in frequency. **(d)** Fitted result obtained using the power law, $\tau = \tau^*_0 [(T_f - T_g) / T_g]^{-zv}$. (e) Fitted result obtained using the empirical Vogel–Fulcher law, $\omega = \omega_0 \exp[-E_a / k_B(T_f - T_0)]$.

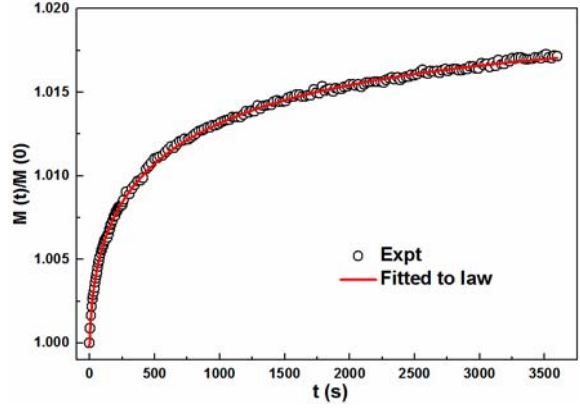

**Figure S2** Time dependence of the normalized magnetization for $(FeTiO_3)_{0.9}$–$(Fe_2O_3)_{0.1}$ measured at 40 K in the ZFC mode. The solid line represents the fit to the stretched exponential expression $M(t)/M(0) = a + b*\exp[-(t/\tau)^\beta]$.

**Table S1.** Fitting parameters for $(FeTiO_3)_{0.9}$–$(Fe_2O_3)_{0.1}$.

| $P$ | $\omega_0$ (rad/s) | $E_a/k_B$ (K) | $T_0$ (K) | $\tau^*_0$ (s) | $T_g$ (K) | $zv$ |
|---|---|---|---|---|---|---|
| 0.02 | $1.26 \times 10^{11}$ | 166 | 37 | $5.81 \times 10^{-11}$ | 45 | 4.19 |

## 2) Verification of the Morin temperature ($T_M$) using a Clausius-Clapeyron type equation

A method to determine the Morin temperature ($T_M$) has been reported recently [Appl. Phys. Lett. 100, 063102 (2012)]. The magnetic transition at $T_M$ satisfies a Clausius-Clapeyron type equation of the form

$$\frac{dT}{dH} = -\mu_0 \frac{\Delta M}{\Delta S}, \tag{S1}$$

where $\Delta M$ is the difference in the magnetization during the transition and $\Delta S$ is the entropy change. **Figure** S3 shows the temperature dependence of the magnetization for ($\alpha$-$Fe_2O_3$)$_{0.1}$–$(FeTiO_3)_{0.9}$ (HI-9) with different cooling fields. When the cooling field ($H_{cool}$) is 0 Oe, the magnetization begins to increase significantly around the characteristic temperature ($T_m = 30$ K). (Explicitly, we chose $T_m$ as the temperature at the maximum of the second derivative of the magnetization.) With increasing cooling field, this transition occurs at lower temperatures (see the black arrow in **Figure** S3). The cooling field dependence of $T_m$ is shown in **Figure** S4. As shown in **Figure** S4, $T_m$ decreases linearly with increasing cooling field and the slope in this case is $dT_m / dH = -0.61$ K/kOe.

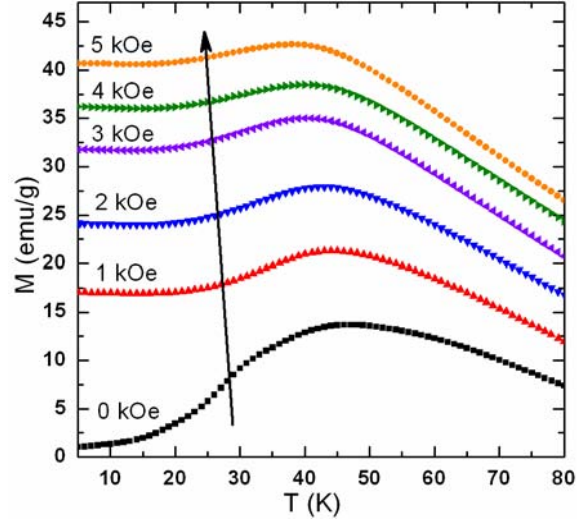

**Figure S3** Temperature dependence of magnetization for $(\alpha\text{-Fe}_2\text{O}_3)_{0.1}$–$(\text{FeTiO}_3)_{0.9}$ with different cooling fields. The arrow shows the tendency of the transition temperature to decrease with increasing cooling field.

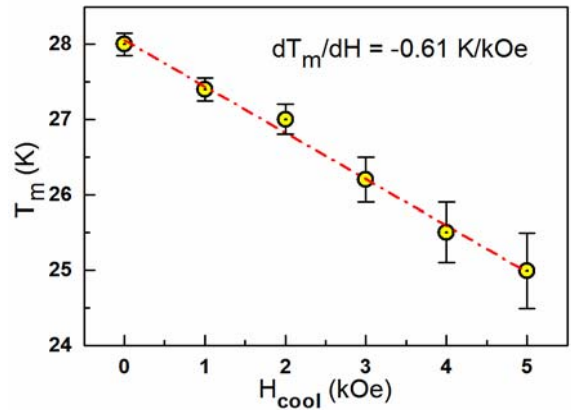

**Figure S4** Cooling field dependence of the characteristic temperature ($T_m$) for $(\alpha\text{-Fe}_2\text{O}_3)_{0.1}$–$\text{FeTiO}_3)_{0.9}$. Circles correspond to the experimental data and the dashed line to a linear fit.

**Figure** S5 shows $\Delta S$ at the maximum applied field ($H = 5$ kOe) near the transition temperature, $T_m$, for $(\alpha\text{-Fe}_2\text{O}_3)_{0.1} - (\text{FeTiO}_3)_{0.9}$. According to Eq. (S1) of the Supplemental Information, using the values determined at $H = 5$ kOe for $\Delta M = 2.05$ emu/g = 2.05 Am$^2$/kg and $\Delta S = 0.33$ J/kgK, we obtain $dT_m / dH = -0.62$ K/kOe. This result agrees well with the measured field dependence of the transition temperature (–0.61 K/kOe) and we identify the transition temperature $T_m$ (~ 30 K) as the Morin temperature $T_M$.

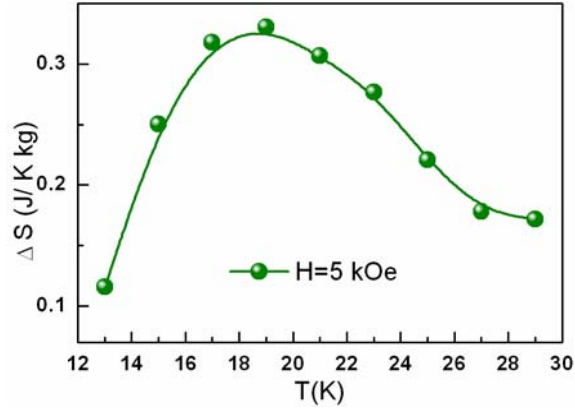

**Figure S5** Entropy change $\Delta S$ at the maximum applied field (5 kOe) at the transition temperature, $T_m$, for $(\alpha\text{-Fe}_2\text{O}_3)_{0.1}$–$(\text{FeTiO}_3)_{0.9}$. The solid line is to guide the eye.

## 3) Revised NS (RNS) model for HI-9

The details of the revised NS model, based on the TPPOPP structure in HI-9, are described in what follows. Because of the interface exchange interaction, $H_E$ can be regarded as arising from the balance between the interface pinning energy, the anisotropy energy and the Zeeman energy of the system, [Phys. Rev. 102, 1413 (1956); Phys. Rev. B 72, 174422 (2005)] which can be simplified as

$$- H_E \propto J_i m_i. \tag{S2}$$

Here, $J_i$ is the interface exchange constant and $m_i$ is the interface moment. For the HI-9 system, $m_i$ arises from two sources: the first part is the field-induced $m_i$ that arises during the dynamic transition from the AFM $Fe^{2+}$ to the FM $Fe^{2+}$ and the second part is the actionless $m_i$.

For the AFM domains (including $Fe^{2+}$ and $Fe^{3+}$), the interface spins can switch, driven by $H_{cool}$. Let $\upsilon_+$ and $\upsilon_-$ denote the switching rates for the A and B sublattices, respectively (A and B denote the two sublattices with opposite spin orientation in the AFM domains). The number of times the interface spins switch during a typical measurement is then $\upsilon_+\tau$ (or $\upsilon_-\tau$), [Phys. Rev. B 72, 174422 (2005); R. C. O'Handley, *Modern Magnetic Materials: Principles and Applications* (John Wiley and Sons, New York, 1999)] where $\tau$ is the typical measurement time (~ $10^2 - 10^3$ s). Thus we can calculate the induced magnetization after the FC process as

$$\frac{\mu_0 g^2 \mu_B^2}{k_B T} H_{cool} = (\nu_+ - \nu_-)\tau. \tag{S3}$$

The first effective magnetic field ($H_{ef1}$) results in the activation energy for the switching process, i.e., $\nu_\pm = \nu_0 e^{-E_\pm/k_B T}$, with $E_\pm = KV \pm \mu H_{ef1}$. Here, $K$ is the anisotropy constant, $V$ is the volume of the AFM domains, and $\nu_0 = 10^9$ s$^{-1}$ is the frequency factor. [R. C. O'Handley, *Modern Magnetic Materials: Principles and Applications* (John Wiley and Sons, New York, 1999)] Using these values for $\nu_\pm$, Eq. (S3) can be rewritten as

$$\frac{\mu_0 g^2 \mu_B^2}{k_B T} H_{cool} \approx 2\nu_0 \tau e^{-KV/k_B T} \sinh(-\mu H_{ef1}). \tag{S4}$$

We can solve for $H_{ef1}$ from Eq. (S4), and obtain

$$H_{ef1} \approx -\text{arcsinh}\left(\frac{\mu_0 g^2 \mu_B^2 e^{KV/k_B T}}{2 k_B T \nu_0 \tau} H_{cool}\right). \tag{S5}$$

For a small cooling field ($\mu_0 H_{cool} < k_B T$), Eq. (S5) can be rewritten as

$$H_{ef1} \propto -\ln(H_{cool}), \tag{S6}$$

where the minus sign indicates that the coupling of $m_i$ to the AFM spins dominates over the coupling of $m_i$ to $H_{cool}$.

For the FM domains (FM Fe$^{2+}$), the coupling energy of $m_i$ to the FM domains (magnetic anisotropy energy) competes with the coupling energy of $m_i$ to $H_{cool}$ (Zeeman energy). These competing energies can be described as the second effective magnetic field ($H_{ef2}$) acting on $m_i$ during the cooling process [Phys. Rev. B 72, 174422 (2005)] ($\mu H_{cool} < k_B T_f$) and takes the form

$$H_{ef2} = \frac{J_i \mu_0}{(g\mu_B)^2} L\left(\frac{\mu H_{cool}}{k_B T_f}\right) + H_{cool}, \tag{S7}$$

where the $g$ is the Lande factor, $\mu_B$ is the Bohr magneton, $L(x)$ is the Langevin function, $\mu = N_v \mu_0$, $N_v$ is the number of the FM spins, and $k_B$ is the Boltzmann constant.

The total effective magnetic field ($H_{ef}$) can be thought of as the sum of $H_{ef1}$ and $H_{ef2}$ such that

$$H_{ef} = H_{ef1} + H_{ef2}. \tag{S8}$$

When cooling to lower temperatures (5 K), because of the SG behavior, the frozen $m_i$ ($\propto H_{ef}$) induces the EB behavior. [Phys. Rev. B 72, 174422 (2005)] Using $H_E \propto J_i m_i$ and noting that $m_i \propto H_{ef}$, we can combine Eqs. (S2) - (S8) and summarize the relationship as

$$-H_E \propto J_i \left[ \frac{J_i \mu_0}{(g\mu_B)^2} L\left(\frac{\mu H_{cool}}{k_B T_f}\right) + H_{cool} - \ln(H_{cool}) \right] \quad (S9)$$

**4) Initial magnetization curves at different temperatures for HI-9**

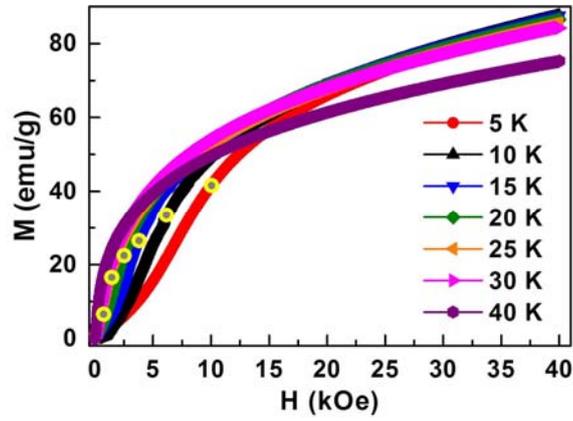

**Figure S6** Initial magnetization curves at different temperatures for HI-9. Gray points with yellow bright border indicate the positions of the critical magnetic field ($H_{cmf}$).